\documentclass[conference]{IEEEtran}
\IEEEoverridecommandlockouts
\usepackage{cite}
\usepackage{amsmath,amssymb,amsfonts}
\usepackage{algorithmic}
\usepackage{graphicx}
\usepackage{textcomp}
\usepackage{xcolor}
\usepackage{epstopdf}
\def\BibTeX{{\rm B\kern-.05em{\sc i\kern-.025em b}\kern-.08em
    T\kern-.1667em\lower.7ex\hbox{E}\kern-.125emX}}
\begin{document}

\title{Deep Learning Based OFDM Channel Estimation Using Frequency-Time Division and Attention Mechanism}

\author{\IEEEauthorblockN{Ang Yang, Peng Sun, Tamrakar Rakesh, Bule Sun, Fei Qin}
\IEEEauthorblockA{Communication Research Institute, vivo Mobile Communication Co., Ltd., Beijing, China \\
Email: \{ang.yang, sunpeng, rakesh, bule.sun, qinfei\}@vivo.com}
}

\maketitle

\begin{abstract}
In this paper, we propose a frequency-time division network (FreqTimeNet) to improve the performance of deep learning (DL) based OFDM channel estimation. This FreqTimeNet is designed based on the orthogonality between the frequency domain and the time domain. In FreqTimeNet, the input is processed by parallel frequency blocks and parallel time blocks sequentially. By introducing the attention mechanism using the SNR information, an attention based FreqTimeNet (AttenFreqTimeNet) is proposed. Using 3rd Generation Partnership Project (3GPP) channel models, the mean square
error (MSE) performance of FreqTimeNet and AttenFreqTimeNet under different scenarios is evaluated. A method for constructing mixed training data is proposed, which could address the generalization problem in DL. It is observed that AttenFreqTimeNet outperforms FreqTimeNet, and FreqTimeNet outperforms other DL networks with reasonable complexity.
\end{abstract}


\section{Introduction}
Orthogonal frequency division multiplexing (OFDM) is one of the core technologies in both 4G and 5G standard, which supports multiple access well and performs robustly in frequency selective fading environment. Moreover, it seems that OFDM is very likely to be reused in 6G standard. In wireless communication systems, the transmit signals suffer from various kinds of fading and multi-path propagations. To demodulate the transmit signals, pilot signals such as demodulation reference signal (DMRS) are designed to estimate the channel information. The DMRS is transmitted along with the data signals, using the same transmit precoding of the data signals and suffering from similar channel fadings. Since the locations and the sequences of DMRS are known to the receiver, the receiver could estimate the channel using the received signals. Least square (LS) and linear minimum mean square error (LMMSE) are two representative OFDM channel estimation methods.

In recent years, deep learning (DL) or artificial intelligence (AI) has been widely investigated in wireless communication systems, in both academia and industries \cite{DL-1,DL-2,Jinshi-AI-Compress-1}. DL is proved to work successfully in various areas, such as MIMO detection \cite{Jinshi-ModelDriven-3}, channel state information (CSI) feedback \cite{Jinshi-CSI-1,Jinshi-CSI-2,Jinshi-CSI-3}, signal recovery \cite{Weichen-1} and channel estimation \cite{Estimation-1,Estimation-2,ChannelNet,ReEsNet}. An end-to-end DL network is proposed in \cite{Estimation-1} to prove the feasibility of DL based OFDM channel estimation. As the combination of the super-resolution network (SRCNN) and the denoising neural network (DnCNN), ChannelNet is proposed to improve the OFDM channel estimation performance \cite{ChannelNet}. Employing Residual learning, which is a powerful tool in image super-resolution, the deep residual channel estimation network (ReEsNet) is proposed in \cite{ReEsNet}.

However, in many DL-based communication studies, DL technologies in computer science are directly applied to wireless communication, ignoring many essential characteristics of communication. Actually, the features of wireless communication channels and the theories of wireless communication are very helpful in the design of DL networks in \cite{Jinshi-ModelDriven-1,Jinshi-ModelDriven-2,Jinshi-ModelDriven-3}.

In this paper, we propose a frequency-time division network (FreqTimeNet) for OFDM channel estimation. The orthogonality between the frequency domain and the time domain, which is used in channel estimation methods in industry to reduce the complexity of the OFDM channel estimator, is the key idea of FreqTimeNet. The input is divided into subsets along the time domain and each subset goes into one of the frequency blocks. Then the outputs of the frequency blocks are combined and then divided again into subsets along the frequency domain. Each new subset passes through one of the time blocks and then the final outputs are obtained by using frequency combination. Furthermore, to involve signal to noise ratio (SNR) in FreqTimeNet, an attention based FreqTimeNet (AttenFreqTimeNet) is proposed to improve the performance under different SNRs. The simulation results are provided under 3rd Generation Partnership Project (3GPP) channel models. In the simulation, we construct a mixed channel model, concluding non-line-of-sight (NLOS) channel with low speed, NLOS channel with high speed, line-of-sight (LOS) channel with low speed and LOS channel with high speed. This mixed training data could reduce the impact of the generalization problem in DL. Better mean square error (MSE) performance is achieved by FreqTimeNet compared to other DL based method and AttenFreqTimeNet outperforms FreqTimeNet.

\section{SYSTEM MODEL}
\begin{figure}[!t]
\centering
\includegraphics[width=3.5in]{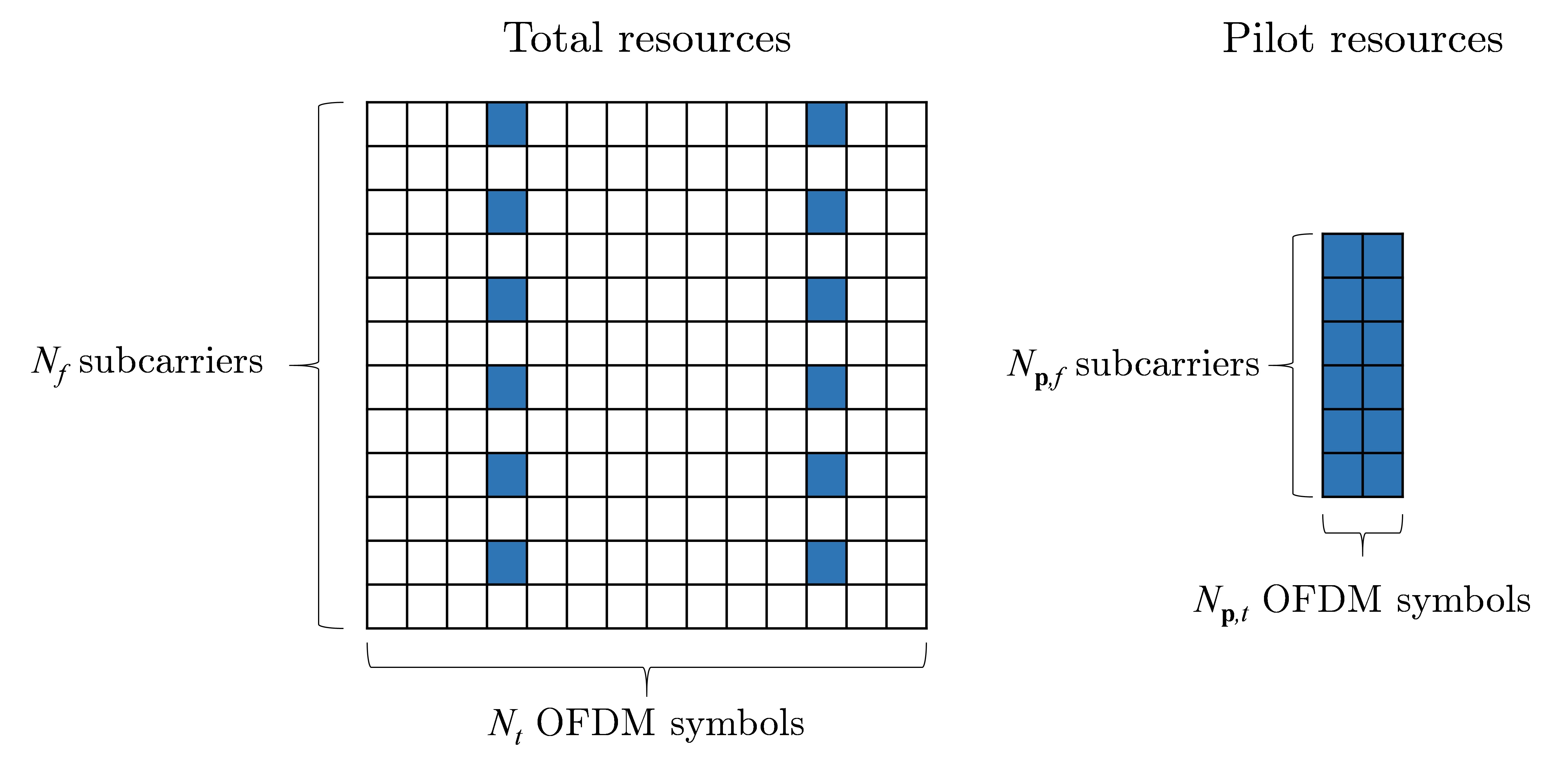}
\caption{An example of pilots in 5G OFDM system.} \label{SystemFig}
\end{figure}

We consider an OFDM wireless communication system with one typical setting in \cite{38.211}. The minimum unit of the resources in time domain is one OFDM symbol, and the minimum unit of the resources in frequency domain is one subcarrier. Resource element (RE), which occupies one OFDM symbol and one subcarrier, is noted as the minimum time-frequency resource. As shown in Fig. \ref{SystemFig}, to assist the demodulation of the received signals, pilots are located sparsely in the time-frequency resources, the rest of which could be used for data transmission and other kinds of reference signals. Fig. \ref{SystemFig} shows one typical pilot resource allocation in current 5G system \cite{38.211}. In the frequency domain, there are 12 subcarriers in one resource block (RB) and 14 OFDM symbols in one time slot. The pilots occupy 6 subcarriers and 2 OFDM symbols in one RB and one time slot.

Assume the system has $N_{t}$ OFDM symbols, $N_{f}$ subcarriers, $M_{Tx}$ transmit antennas and $M_{Rx}$ receive antennas. For the $k$th OFDM symbol, the $i$th subcarriers and the $m_{Rx}$th receive antenna, the received signal at the receiver can be represented by
\begin{equation}\label{receiver 1}
y_{k,i,m_{Rx}} = {\sum_{m_{Tx}=1}^{M_{Tx}} h_{k,i,m_{Tx},m_{Rx}}} w_{k,i,m_{Tx}} s_{k,i} + z_{k,i,m_{Rx}}
\end{equation}where $h_{k,i,m_{Tx},m_{Rx}}$ and $w_{k,i,m_{Tx}}$ are the channel and transmit percoder of the $m_{Tx}$th transmit antenna, respectively. Symbol $s_{k,i}$ denotes the transmit signal and $z_{k,i,m_{Rx}}$ is the white Gaussian noise (AWGN).

Since the pilots and the data signals are beamformed with the same transmit precoders, the above equation could be rewritten as
\begin{equation}\label{receiver 2}
y_{k,i,m_{Rx}} = {\tilde h}_{k,i,m_{Rx}} s_{k,i} + z_{k,i,m_{Rx}}
\end{equation}where ${\tilde h}_{k,i,m_{Rx}} = {\sum_{m_{Tx}=1}^{M_{Tx}} h_{k,i,m_{Tx},m_{Rx}}} w_{k,i,m_{Tx}}$ is the precoded channel. Then for all OFDM symbols and subcarriers, we have
\begin{equation}\label{receiver 3}
{\bf {Y}}_{m_{Rx}} = {\tilde {\bf {H}}}_{m_{Rx}} 	\circ {\bf s} + {\bf z}_{m_{Rx}}
\end{equation}where $y_{k,i,m_{Rx}}$, ${\tilde h}_{k,i,m_{Rx}}$, $s_{k,i}$ and $z_{k,i,m_{Rx}}$ are $\left( k,i \right)$th element of ${\bf {Y}}_{m_{Rx}}$, ${\tilde {\bf {H}}}_{m_{Rx}}$, ${\bf s}$, and ${\bf z}_{m_{Rx}} \in \mathbb{C} ^{N_{f},N_{t}}$, respectively. Symbol $\circ $ denotes the element-wise product, which is also known as the Hadamard product.

Refering to Fig. \ref{SystemFig}, pilots occupies $N_{{\bf p},t}$ OFDM symbols of total $N_{t}$ OFDM symbol and $N_{{\bf p},t}$ subcarriers of total $N_{f}$ subcarriers, with $N_{{\bf p},t} < N_{t},N_{{\bf p},f} < N_{f}$. Focusing on the time-frequency resources occupied by the pilots, for the $m_{Rx}$th receive antenna, we have
\begin{equation}\label{receiver pilot 1}
{\bf {Y}}_{{\bf p},m_{Rx}} = {\tilde {\bf {H}}}_{{\bf p},m_{Rx}} 	\circ {\bf s}_{\bf p} + {\bf z}_{{\bf p},m_{Rx}}
\end{equation} where ${\bf Y}_{{\bf p},m_{Rx}}\in \mathbb{C} ^{N_{{\bf p},t},N_{{\bf p},f}}$ is the received signal on the pilot resources. The channel coefficients of the pilot resources are ${\tilde {\bf {H}}}_{{\bf p},m_{Rx}}$, the pilot signals are ${\bf s}_{\bf p}$, and the AWGN of the pilot resources are ${\bf z}_{{\bf p},m_{Rx}}$. It is clear that ${\bf Y}_{{\bf p},m_{Rx}}$, ${\tilde {\bf {H}}}_{{\bf p},m_{Rx}} $, ${\bf s}_{\bf p} $ and ${\bf z}_{{\bf p},m_{Rx}}$ are subsets of ${\bf Y}_{m_{Rx}}$, ${\tilde {\bf {H}}}_{m_{Rx}} $, ${\bf s}$ and ${\bf z}_{m_{Rx}}$, respectively.

Since ${\bf s}_{\bf p}$ is known by both the transmitter and the receiver, ${\tilde {\bf {H}}}_{{\bf p},m_{Rx}}$ could be estimated based on ${\bf Y}_{{\bf p},m_{Rx}}$ and ${\bf s}_{\bf p}$. Then with the estimation of ${\tilde {\bf {H}}}_{{\bf p},m_{Rx}}$, ${\tilde {\bf {H}}}_{m_{Rx}}$ could be further estimated and used for demodulations of data signals. There are several conventional methods for this problem, such as least square (LS) method and minimum mean square error (MMSE) method \cite{Estimation-1}.

\section{DEEP LEARNING BASED CHANNEL ESTIMATION METHOD}

In this paper, we focus on the problem of estimating ${\tilde {\bf {H}}}_{m_{Rx}}$ based on ${\tilde {\bf {H}}}_{{\bf p},m_{Rx}}$. It can be seen from Fig. \ref{SystemFig} that this problem is similar to the image super-resolution (SR) problem in computer vision. In this classic computer vision problem, a low-resolution image with or without noise is processed to a high-resolution image with the best possible image quality.

Regarding our problem, ${\tilde {\bf {H}}}_{{\bf p},m_{Rx}}$ could be treated as the low-resolution image, the size of which is $N_{{\bf p},t} \times N_{{\bf p},f} \times 2$, and ${\tilde {\bf {H}}}_{m_{Rx}}$ could be seen as the high-resolution image, the size of which is $N_{t} \times N_{f} \times 2$. Based on this logic, deep learning, which is one powerful tool in image super-resolution, could be used to solve the problem.

\begin{figure*}[!t]
\centering
\includegraphics[width=4.8in]{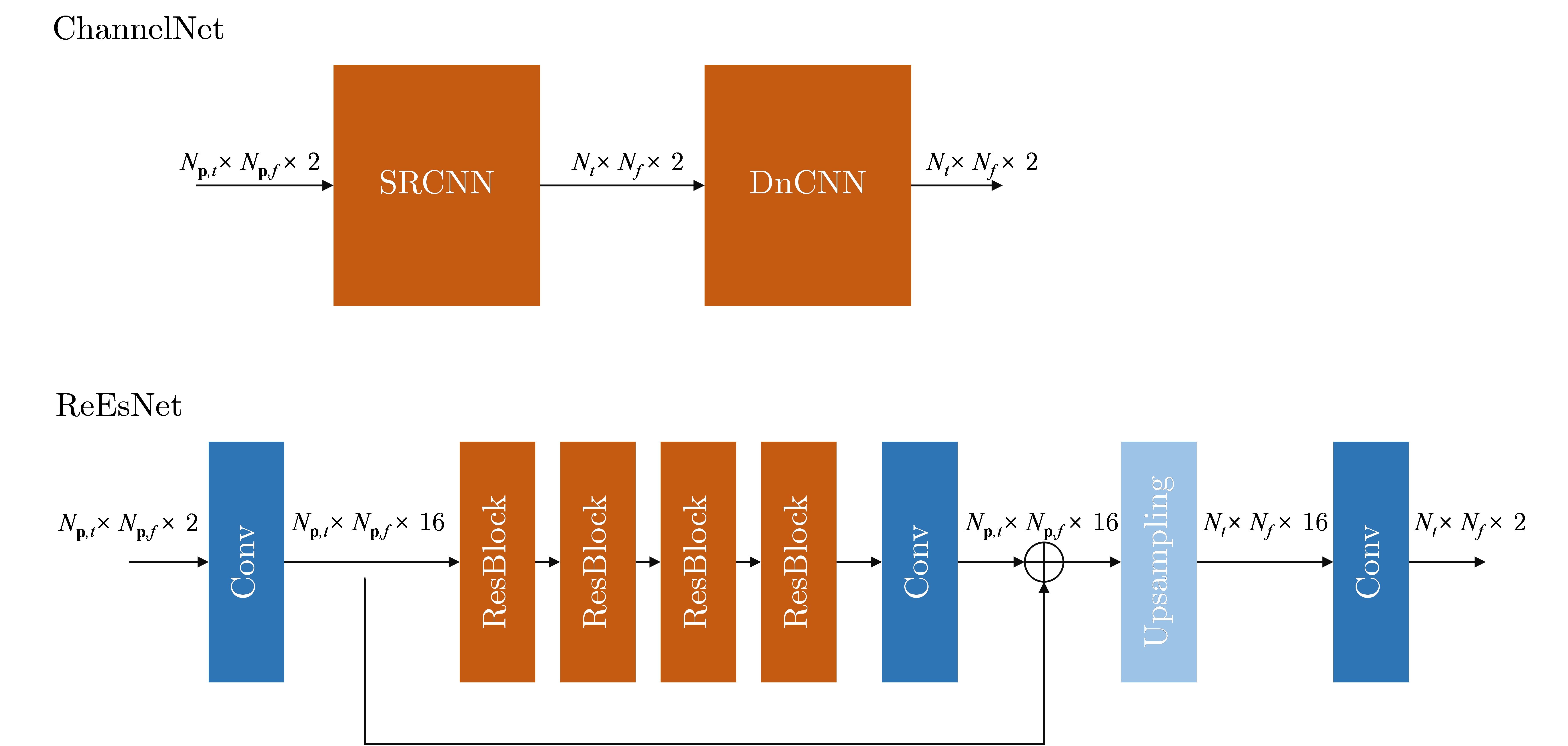}
\caption{Architectures of ChannelNet and ReEsNet.} \label{Net1}
\end{figure*}

\subsection{ChannelNet}
ChannelNet is the combination of two neural networks \cite{ChannelNet}, as shown in Fig. \ref{Net1}. The first neural network is SRCNN, which is used to transform the low-resolution image of size $N_{{\bf p},t} \times N_{{\bf p},f} \times 2$ to the high-resolution image with size $N_{t} \times N_{f} \times 2$. The second neural network is DnCNN, which is used for noise reduction and does not change the size of the image. Note that in \cite{ChannelNet}, only one
filter of size $5 \times 5$ is used in the last layer of SRCNN and the output of SRCNN would be $N_{t} \times N_{f} \times 1$. Here two filters of size $5 \times 5$ are deployed to improve the performance.

\subsection{ReEsNet}
ReEsNet is based on residual learning, which is introduced to solve the gradient vanishing problem and the gradient explosion problem in very deep DL network \cite{ReEsNet}, as shown in Fig. \ref{Net1}. With the combination of identity mapping and residual mapping, these gradient related problems could be mitigated. After one convolutional layer with 16 filters of size $3 \times 3 \times 2$, the input of size $N_{{\bf p},t} \times N_{{\bf p},f} \times 2$ is transformed to the feature map of size $N_{{\bf p},t} \times N_{{\bf p},f} \times 16$. After 4 ResBlocks and one convolutional layer, the size of the feature map remains the same. The transposed convolution is used for up-sampling. After up-sampling, the size of the feature map is increased to $N_{t} \times N_{f} \times 16$. The final output is obtained after another convolutional layer.

\subsection{FreqTimeNet}

\begin{figure*}[!t]
\centering
\includegraphics[width=6.1in]{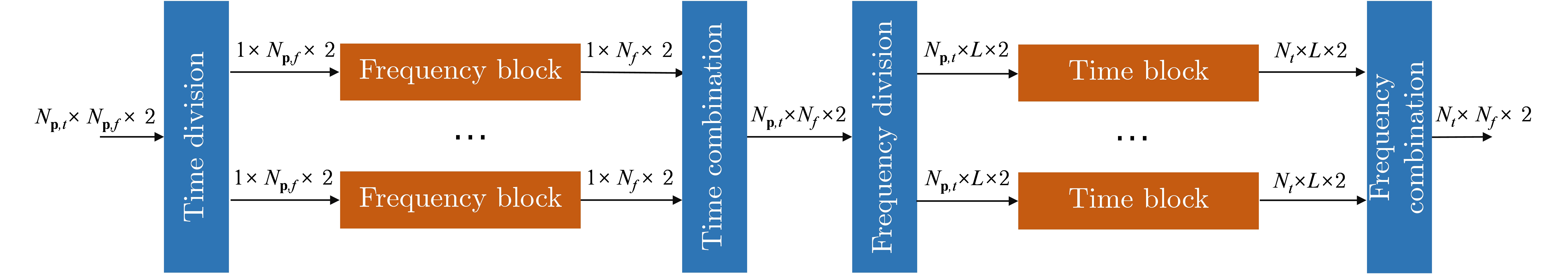}
\caption{Architecture of FreqTimeNet.} \label{Net2}
\end{figure*}

From Eq. (\ref{receiver pilot 1}), it is seen that ${\tilde {\bf {H}}}_{{\bf p},m_{Rx}}$ is 2D matrix, where one dimension is frequency and the other dimension is time. Then the channel estimation methods usually employ 2D matrix operations or 2D integral. For current wireless communication system with large bandwidth in 5G, there might be thousands of subcarriers, while there are 12 or 14 OFDM symbols in one time slot \cite{38.211}. The 2D matrix operations or 2D integral would be too complicated to be used in practical wireless communication systems. To our knowledge, to reduce the complexity, one commonly used channel estimation method in industry is based on the orthogonality between the frequency domain and the time domain. That is, in the first step, the channel resources are divided in time domain, and on each OFDM symbol, the channel coefficients of all subcarriers are recovered first based on the subcarriers of the pilots, separately. In the second step, the channel resources are divided in frequency domain, and on each small number of subcarriers, the channel coefficients of all OFDM symbols are recovered based on the OFDM symbols of the pilots, separately. Based on this frequency-time division method, similar performance is achieved with much lower complexity.

Using the principle of this frequency-time division method, FreqTimeNet is proposed in this paper for the OFDM channel estimation. The architecture of FreqTimeNet is presented in Fig. \ref{Net2}. The size of the input is $N_{{\bf p},t} \times N_{{\bf p},f} \times 2$. In the time division module, the input is divided into $N_{{\bf p},t}$ parts, and the size of each part is $1 \times N_{{\bf p},f} \times 2$. Each part is reshaped into a vector and then goes through one of the frequency blocks, which is a small full-connected (FC) network with one hidden layer having $N_{{\bf p},f} \times 3$ neurons. The outputs of the frequency blocks are reshaped into the $1 \times N_{f} \times 2$ feature maps and are combined as one $N_{{\bf p},t} \times N_{f} \times 2$ feature map in time dimension. In the following, the $N_{{\bf p},t} \times N_{f} \times 2$ feature map is divided into $\frac{N_{f} }{L}$ parts, and the size of each part is $N_{{\bf p},t} \times L \times 2$. Then each part is reshaped into a vector and  then goes through one of the time blocks, which is a small full-connected network with one hidden layer having $N_{{\bf p},t} \times L \times 2$ neurons. The outputs of the frequency blocks are reshaped into the $N_{t} \times L \times 2$ feature map. After combining all the feature maps, the final output of size $N_{t} \times N_{f} \times 2$ could be obtained. Note that rectified linear unit (ReLu) is used as the activation function of the FC networks.

All the frequency blocks could use the same parameters, and all the time blocks could also share the parameters, which could largely reduce the number of parameters. Note that, we use simple full-connected network in both frequency blocks and time blocks, which could be further optimized by using convolutional neural network (CNN).

\subsection{AttenFreqTimeNet}

\begin{figure*}[!t]
\centering
\includegraphics[width=6.8in]{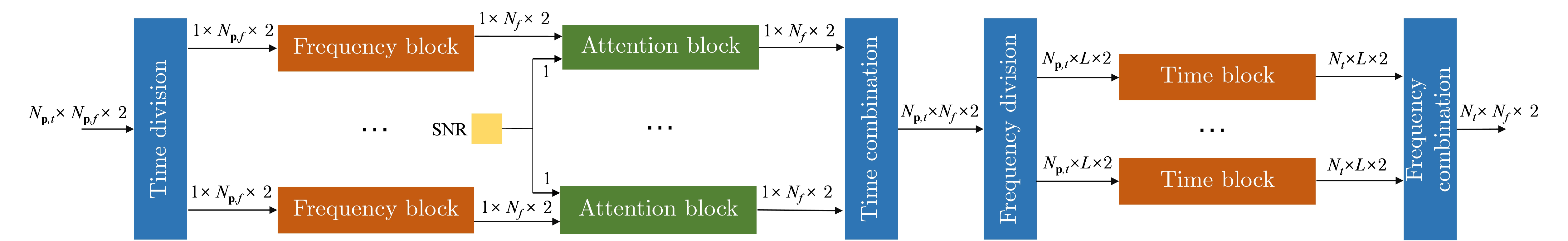}
\caption{Architecture of AttenFreqTimeNet.} \label{Net3}
\end{figure*}

\begin{figure*}[!t]
\centering
\includegraphics[width=5in]{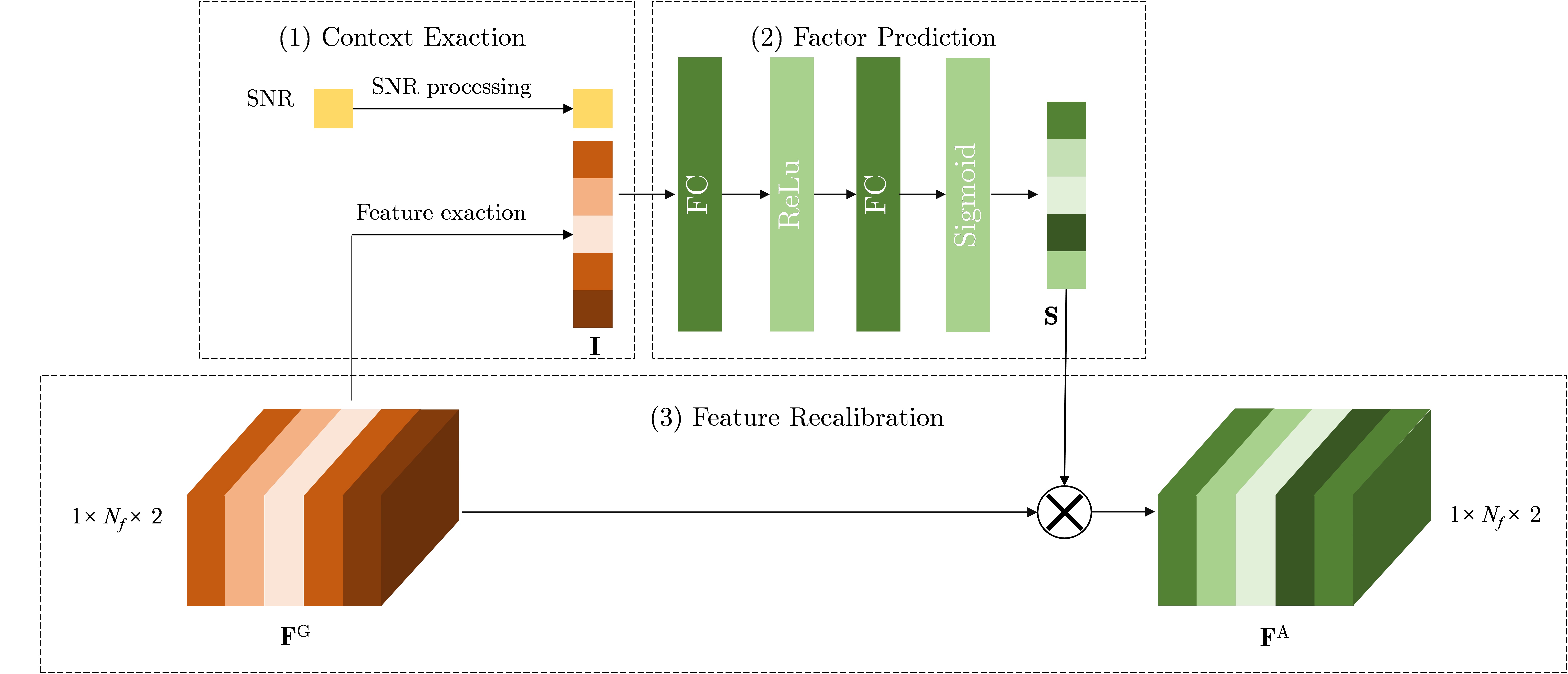}
\caption{Architecture of the attention block.} \label{Net4}
\end{figure*}

Attention mechanism is a DL technique widely used in natural language processing and computer vision \cite{Attention-1}. Recently, attention mechanism has also been employed in wireless communications \cite{Attention-2,Attention-3}. This mechanism introduces additional neural network, which can select different features in the original neural network according to different situations. Also, this additional neural network could assign different weights to the original features and these weights could be called as the soft attentions. After this process, the performance of the neural network could be improved, especially for the data under various situations.

SNR is a very important channel state information and could be easily acquired in current wireless networks \cite{38.214}. However, SNR is not used in above neural networks. How to design the neural network with SNR is a topic deserving research. Then attention mechanism is introduced to involve SNR in the FreqTimeNet, and the new neural network is called as AttenFreqTimeNet, which is shown in Fig. \ref{Net3}. After each frequency block, one attention block is added to pay attention to different features in different SNRs. The details of the attention block could be seen in Fig. \ref{Net4}. Note that The output of one frequency block is ${\bf{F}}^{G}$ with size $1 \times N_{f} \times 2$.

An attention block includes three parts: 1) context extraction; 2) factor prediction; and 3) feature recalibration.

{\emph { 1) Context extraction: }} The context information $\bf{I}$ includes two parts. The first part is the SNR related information and the second part is the output of the feature exaction of ${\bf{F}}^{G}$. In this paper, The SNR related information is obtained through a simple FC network, in which the input is the linear value of SNR, the hidden layer has 50 neurons, and the output layer has 10 neurons. Linear method is employed for feature exaction and then ${\bf{F}}^{G}$ is directly used as a part of the context information $\bf{I}$.

{\emph { 2) Factor prediction: }} A factor prediction neural network is employed to obtain the scaling factor $S$ under different SNRs. Here a simple neural network with two FC layers is used. The first FC layer has $1 \times N_{f}$ neurons with a ReLu. The second FC layer has $1 \times N_{f} \times 2$ neurons with a Sigmoid, which could limit the output range to (0,1) and achieve better performance than ReLu.

{\emph { 3) Feature recalibration: }} The recalibrated feature map ${\bf{F}}^{A}$ is obtained by the element-wise product of ${\bf{F}}^{G}$ and the scaling factor $S$. The impacts of different SNRs have been included in ${\bf{F}}^{A}$.

\section{SIMULATION RESULTS}
In this section, we present numerical results of the noted networks based on link level simulations. We consider single transmit antenna and single receive antenna. There are $N_{f} = 96$ subcarriers in the frequency domain and $N_{t} = 14$ OFDM symbols in the time domain. This is equivalent to 8 RBs in frequency domain since there are 12 subcarriers in one RB, and one time slot in time domain. The pilot pattern as depicted in Fig. \ref{SystemFig} is used. The number of pilots is 96, and in other words, the pilots occupy total 96 resource elements. The link level simulator follows 3GPP tapped delay line (TDL) models \cite{39.901}, which has been calibrated. The carrier frequency is 3.5GHz, and subcarrier space is 15KHz. A new method for constructing mixed training data is proposed to address the generalization problem in DL. For the training data, the channel model is a mixed model of TDL-A, TDL-B, TDL-C, TDL-D and TDL-E, where one sample randomly selects one channel model from these 5 models; the delay spread is randomly chosen from 0ns to 300ns; the speed is randomly generated from 0km/h to 50km/h; the SNR of each sample is randomly selected from [0dB, 5dB, 10dB, 15dB, 20dB]. We use total 90,000 training samples, 10,000 validation samples and 10,000 test samples. That is, 2000 test samples are used for one SNR value in one simulation figure. The number of total epochs is 100, the size of mini-batch is 128, and the optimizer is Adam with the default setting in Keras. The MSE between the output of DL network and actual channel information is used for both training and performance evaluation.

During the training of FreqTimeNet and AttenFreqTimeNet, all the time blocks share their parameters, but the frequency blocks use separate parameters, as a trade-off between performance and complexity. The attention blocks also use separate parameters. The hype-parameter $L$ is set as 12 and then each time block deals with 12 subcarriers, which is 1 RB in the frequency domain. ReEsNet 1 uses the same hyper-parameters in \cite{ReEsNet}, while in ReEsNet 2, the number of filters in each convolutional layer is 32 except the last convolutional layer and the number of ResBlocks is 6.

\begin{figure}[!t]
\centering
\includegraphics[width=3in]{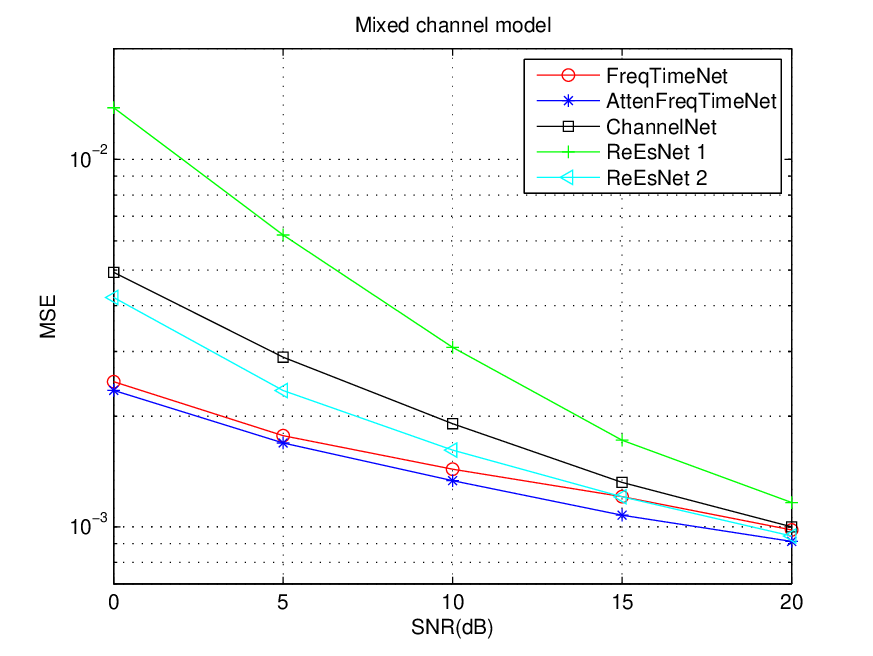}
\caption{The performance of DL networks in mixed channel model.} \label{result1}
\end{figure}

The performance of the proposed FreqTimeNet and AttenFreqTimeNet for the mixed channel model is shown in Fig. \ref{result1}, along with ChannelNet, ReEsNet 1 and ReEsNet 2. The settings of mixed channel model are the same as the training samples. It is seen that AttenFreqTimeNet achieves the best MSE performance in all SNR points, and the performance of ReEsNet 1 is the worst. Note that the same hype-parameters for ReEsNet 1 are used as \cite{ReEsNet}, but the channel model and simulation details are different from \cite{ReEsNet}. Other values of hype-parameters would improve the performance of ReEsNet. Then ReEsNet 2 with higher complexity is investigated. For MSE of $2\times 10^{-3}$, FreqTimeNet achieves about 4dB SNR gain compared to ReEsNet 2, and more gain compared to ChannelNet and ReEsNet 1. As the SNR increases, the advantage of FreqTimeNet diminishes gradually, which means that the generalization performance of FreqTimeNet in high SNR needs to be further improved. Since the attention blocks are used in AttenFreqTimeNet to improve the performance under different SNRs, the advantage of AttenFreqTimeNet is stable in various SNRs.

\begin{figure}[!t]
\centering
\includegraphics[width=3in]{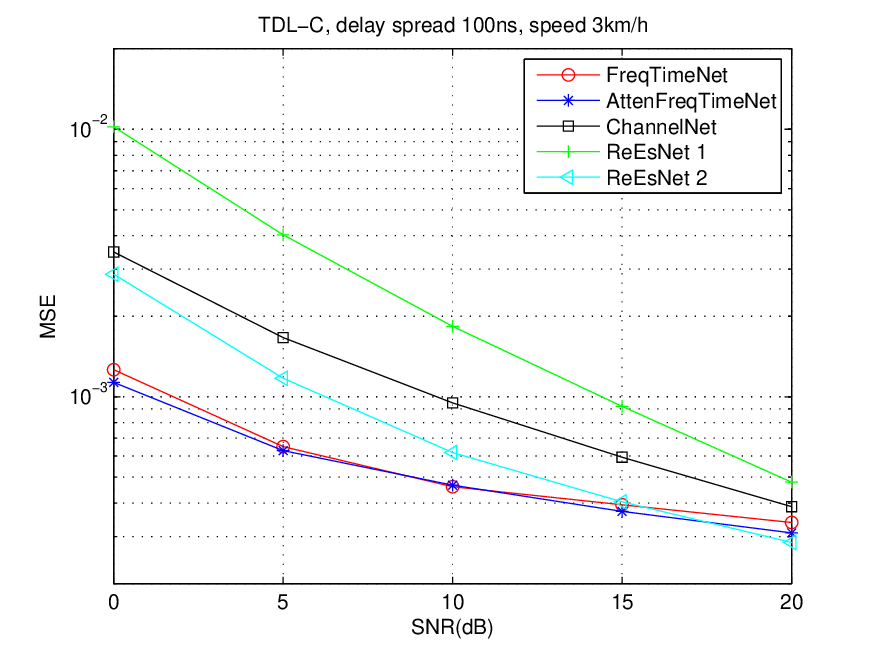}
\caption{The performance of DL networks for TDL-C model, delay spread 100ns, and speed 3km/h.} \label{result2}
\end{figure}

\begin{figure}[!t]
\centering
\includegraphics[width=3in]{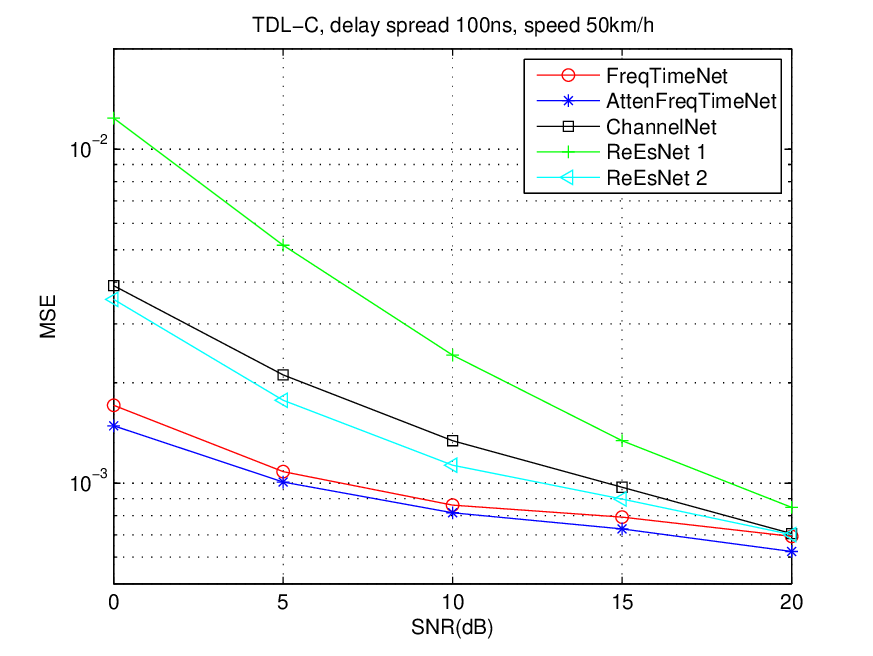}
\caption{The performance of DL networks for TDL-C model, delay spread 100ns, and speed 50km/h.} \label{result3}
\end{figure}

\begin{figure}[!t]
\centering
\includegraphics[width=3in]{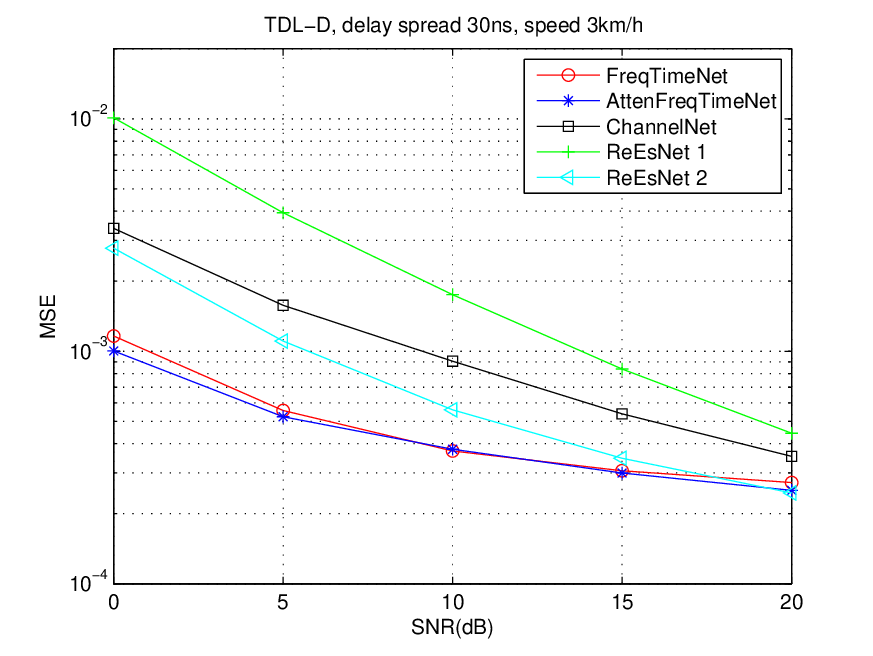}
\caption{The performance of DL networks for TDL-D model, delay spread 30ns, and speed 3km/h.} \label{result4}
\end{figure}

\begin{figure}[!t]
\centering
\includegraphics[width=3in]{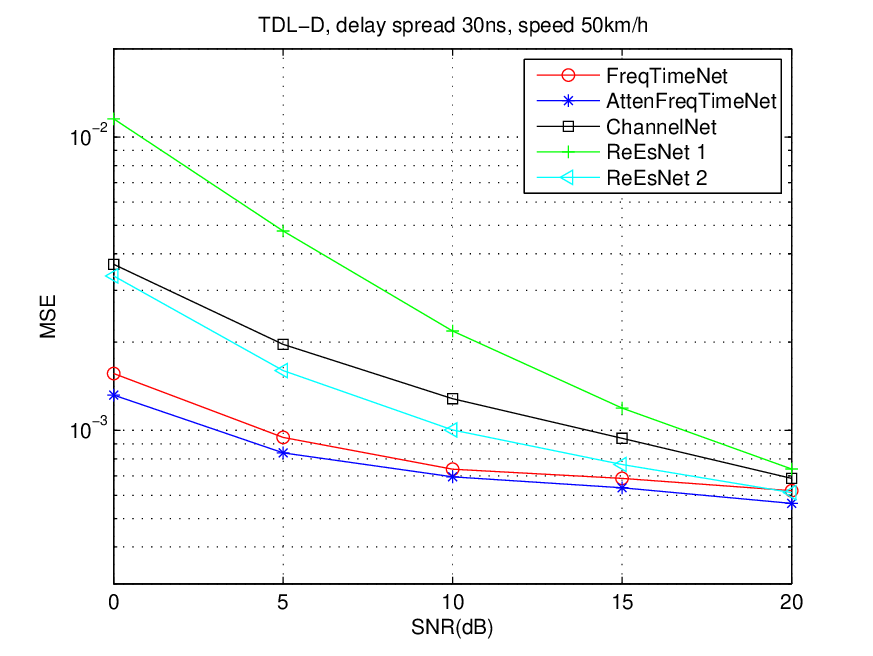}
\caption{The performance of DL networks for TDL-D model, delay spread 30ns, and speed 50km/h.} \label{result5}
\end{figure}

In the following, we test the DL networks in different scenarios. First, we focus on NLOS scenario and use the TDL-C model with delay spread of 100ns. The evaluation results for low speed of 3km/h are shown in Fig. \ref{result2} and results for the high speed of 50km/h are shown in Fig. \ref{result3}. Then, LOS scenario is considered and one typical setting is adopted, in which the model is TDL-D and the delay spread is 30ns. Fig. \ref{result4} and Fig. \ref{result5} show the results of 3km/h and 50km/h in this scenario, respectively. From the curves in these 4 figures, it can be seen that AttenFreqTimeNet achieves almost the best MSE performance in different channel conditions. Moreover, it is seen that these DL networks provide stable performance in different scenarios.

The numbers of the parameters and the flops of the DL networks are presented in Table \ref{tab1}, using the application programming interface (API) of Keras. It is noticed that the complexity of FreqTimeNet is much lower than ChannelNet, higher than ReEsNet 1, and similar to ReEsNet 2. Considering the good MSE performance of FreqTimeNet, the complexity is acceptable. The complexity of AttenFreqTimeNet is higher than FreqTimeNet, since the attention blocks bring extra calculations. Note that the frequency blocks, the time blocks and the attention blocks are small FC networks, and then it is clear that employing convolutional neural network (CNN) could largely reduce the complexity of FreqTimeNet and AttenFreqTimeNet.

\begin{table}[!t]
\caption{The complexity analysis of DL networks}
\begin{center}
\begin{tabular}{|c|c|c|}
\hline
\textbf{Methods} & \textbf{{Number of Parameters}}& \textbf{{Number of Flops}} \\
\hline
FreqTimeNet & 102K & 286k \\
\hline
AttenFreqTimeNet·& 147K & 416k \\
\hline
ChannelNet & 686K & 1364K \\
\hline
ReEsNet 1 & 27K & 54K \\
\hline
ReEsNet 2 & 145K & 289K \\
\hline
\end{tabular}
\label{tab1}
\end{center}
\end{table}

\section{CONCLUSIONS}
In this paper, for DL based OFDM channel estimation, the FreqTimeNet has been proposed, which uses both the communication domain knowledge and the DL domain knowledge. Using the orthogonality between the frequency domain and the time domain, the FreqTimeNet is divided into two parts. The first part is parallel frequency learning and the second part is parallel time learning. Moreover, AttenFreqTimeNet has been proposed to use the SNR information with attention mechanism. The simulation results have been provided under 3GPP channel models. A method for constructing mixed training data has been proposed to deal with the generalization problem in DL. It has been shown that in different communication scenarios, the MSE performance of AttenFreqTimeNet is better than FreqTimeNet and FreqTimeNet achieves lower MSE than other DL networks.

\end{document}